# Size dependence of the properties of synthetic-antiferromagnet-based stochastic magnetic tunnel junctions for probabilistic computing

Takuma Kinoshita[1,2], Ju-Young Yoon[1], Nuno Caçoilo[1], Ryota Mochizuki[1,2], Haruna Kaneko[1,2], Shun Kanai[1-7]*, Hideo Ohno[1,3,4,8], and Shunsuke Fukami[1-4,8,9]

[1]*Laboratory for Nanoelectronics and Spintronics, Research Institute of Electrical Communication, Tohoku University, Sendai 980-8577, Japan*

[2]*Graduate School of Engineering, Tohoku University, Sendai 980-0845, Japan*

[3]*WPI-Advanced Institute for Materials Research, Tohoku University, Sendai 980-8577, Japan*

[4]*Center for Science and Innovation in Spintronics, Tohoku University, Sendai 980-8577, Japan*

[5]*Precursory Research for Embryonic Science and Technology (Japan Science and Technology Agency), Kawaguchi 332-0012, Japan*

[6]*Division for the Establishment of Frontier Sciences, Tohoku University, Sendai 980-8577, Japan*

[7]*National Institutes for Quantum Science and Technology, Takasaki 370-1207, Japan*

[8]*Center for Innovative Integrated Electronic Systems, Tohoku University, Sendai 980-0845, Japan*

[9]*Inamori Research Institute for Science, Kyoto 600-8411, Japan*

*E-mail: skanai@tohoku.ac.jp

Stochastic magnetic tunnel junctions (s-MTJs) are core components for spintronics-based probabilistic computing (p-computing), a promising candidate for energy-efficient unconventional computing. To achieve reliable performance under practical conditions, the use of a synthetic antiferromagnetic (SAF) free-layer configuration was proposed due to its enhanced tolerance to magnetic field perturbations. For engineering the SAF s-MTJs, we systematically investigate the properties of the SAF s-MTJs as a function of the junction size. We observe that decreasing junction size leads to shorter relaxation times, enhanced magnetic field robustness, and enhanced insensitivity to bias voltage. These findings provide key insights toward high-performance p-computers with reliable operation.

Probabilistic computers (p-computers), which utilize the inherent stochasticity in various devices and materials[1–9], have drawn increasing attention due to their significant potential for energy-efficient approaches to complex problems[10]. In particular, stochastic magnetic tunnel junctions (s-MTJs) with a low energy barrier[1,2,11–21] are considered promising due to their ultralow-energy operation without the need for external input to induce stochasticity. Additionally, they are compatible with established design and fabrication technologies of non-volatile magnetoresistive random-access memory (MRAM). Up to now, several proof-of-concepts of p-computers with s-MTJs have been demonstrated for various computationally hard tasks[22–31]. To achieve high computation performance and reliability under practical conditions in p-computing with s-MTJs, it is essential to systematically investigate and identify the factors governing their stochastic behavior, including fluctuation speed and response to external perturbations such as magnetic fields and bias voltages, which largely depend on the configuration of the magnetic layers[14–17,20,21].

In s-MTJs with perpendicular easy axis, the susceptibility to external magnetic fields has been comprehensively investigated, where variations in stochastic behavior have been explained in terms of changes in the Zeeman energy, which scales with the magnetic moment of the free layer ($\propto$ device size)[32]. Recently, it has been reported that the field susceptibility can be drastically suppressed in an in-plane easy-axis s-MTJ with a synthetic antiferromagnetic (SAF) configuration that effectively compensates the Zeeman energy in the free layer[17]. While this in-plane SAF s-MTJ provides a pathway for reliable spintronics-based p-computing, the key factors influencing its stochastic characteristics need to be better understood. For this purpose, in this work we investigate the junction-size dependence of stochastic properties of the SAF s-MTJs under external perturbations, such as magnetic fields and bias voltages.

We prepare a stack of s-MTJs consisting of, from the substrate side, Ta (5.0 nm)/ PtMn

(20 nm)/ Co (2.0 nm)/ Ru (0.85 nm)/ CoFeB (2.2 nm)/ MgO (1.1 nm)/ CoFeB (1.8 nm)/ Ru (0.74 nm)/ CoFeB (2.3 nm)/ Ta (5.0 nm)/ Ru (5.0 nm), deposited on a Si/SiO$_2$ substrate by dc and rf magnetron sputtering at room temperature. The two CoFeB layers above the MgO barrier are antiferromagnetically coupled through the Ru spacer via Ruderman-Kittel-Kasuya-Yosida (RKKY) interaction, forming a SAF-free layer structure (Fig. 1(a)). The nominal thickness of the two CoFeB layers in the SAF-free layer is designed to compensate the magnetic moment of each other, considering the different amounts of magnetically dead layers caused by the different adjacent layers. The stacks are patterned into circular MTJ devices using electron-beam lithography. Figure 1(b) shows a scanning electron microscopy image of a patterned junction, proving a circular shape. After the patterning, the MTJs are annealed at 300°C for two hours under an in-plane magnetic field $\mu_0 H_x = 1.2$ T (where $\mu_0$ is the vacuum permeability), to pin the reference layer magnetization to the in-plane ($+x$) direction. Figure 1(c) shows an $R$-$\mu_0 H_x$ curve, exhibiting superparamagnetic behavior with zero coercive fields. The resistance area ($RA$) product in the antiparallel (AP) and parallel (P) states is determined to be 44±1 and 24±1 Ω·μm$^2$, respectively. The electrically active diameter $D_{\text{ele}}$ of each MTJ is calculated by dividing the $R_\text{P}A$ by the measured device resistance in the P state. The typical tunnel magnetoresistance ratio measured at device level is around 85%.

Figure 1(d) shows a magnetization curve ($M$-$\mu_0 H_x$ curve) of the SAF free-layer stack, which consists of a structure of Ta (5.0 nm)/ PtMn (20 nm)/ Co (0.2 nm)/ Ru (0.85 nm)/ CoFeB (0.3 nm)/ MgO (1.1 nm)/ CoFeB (1.8 nm)/ Ru (0.74 nm)/ CoFeB (2.3 nm)/ Ta (5.0 nm)/ Ru (5.0 nm), measured by vibrating sample magnetometer (VSM). We note that the magnetic layer thicknesses of the reference layer are intentionally reduced so that the magnetic properties of only the free layer are detected. The effective magnetic field of the RKKY interaction ($\mu_0 H_{\text{RKKY}}$) is estimated to be about 80 mT, obtained from the saturation field indicated by the dashed line in Fig. 1(d). While this value reflects the RKKY coupling strength of our stack structure[33,34], it is derived from a blanket film and may differ to some extent after patterning.

The random telegraph noise (RTN) signal is measured using the electrical circuit shown in Fig. 2(a). The magnitude of applied dc is determined such that the averaged voltage across the MTJ becomes 100 mV throughout this work. Figure 2(b) shows a representative RTN signal for a device with $D_{ele}$ = 33 nm, along with a histogram showing the distribution of voltage levels $V$. Fluctuation between two stable states is observed. Note that the TMR ratio deduced from this RTN signal is approximately 23%, which differs from the $R$-$\mu_0 H_x$ curve in Fig. 1(c) with the TMR ratio of around 90%. This discrepancy suggests that the magnetization direction of the free layer in the AP (P) state is not completely antiparallel (parallel) to the reference layer due to the circular shape of the s-MTJs, resulting in the reduced TMR ratio observed in the RTN signal. The energetically stable states in circular MTJs can be formed by strain-induced and/or magnetocrystalline anisotropy, whose direction is randomly oriented[18]. The RTN signal is digitized using a threshold voltage determined by averaging all voltage levels indicated by the horizontal broken line. The event time $t_{event,AP}$ ($t_{event,P}$) refers to the duration that the voltage levels remain continuously above (below) this threshold. Figure 2(c) plots histograms of the number of magnetization reversals $N(t_{event})$, which follow an exponential distribution as indicated by the solid lines: $N(t_{event}) \propto \exp(-t_{event}/\tau)$. The characteristic relaxation time $\tau \equiv \langle t_{event} \rangle$ is derived from these histograms, based on more than 20,000 recorded magnetization reversals.

First, we present the average relaxation time $\tau_{ave} \equiv \sqrt{\tau_P \tau_{AP}}$ of SAF s-MTJs as a function of $D_{ele}$ in Fig. 3(a). $\tau_{ave}$ is measured under a certain $\mu_0 H_x$, adjusted so that $\tau_P = \tau_{AP}$. As $D_{ele}$ decreases, $\tau_{ave}$ reduces significantly from seconds to a few tens of microseconds, favorable for applications in terms of the scalability. We note that the device-to-device variation for junctions with similar $D_{ele}$ may be caused by a variation of strain-induced and/or magnetocrystalline anisotropy[18]. Assuming that the exchange coupling field due to the RKKY interaction is dominant over the anisotropy and external fields, a static-process calculation[17] is applied:

$$\tau_{ave} = \tau_0 \exp \Delta = \tau_0 \exp \frac{\pi \mu_0 H_{K,in} D^2 (m_1 + m_2)}{8 k_B T}, \quad (1)$$

where $\tau_0$, $\Delta$, $H_{K,in}$, $D$, $m_i$, $k_B$, and $T$ are the attempt time, thermal stability factor, effective in-plane magnetic anisotropy field, magnetically active diameter, magnetic moment of the $i$th free layer in the SAF-free layer ($i = 1$: upper free layer; $i = 2$: lower free layer), the Boltzmann constant, and the temperature, respectively. Using the values of $m_i$ determined from the VSM measurements and $T = 300$ K, and assuming $\tau_0 = 1$ ns–1 µs, $\mu_0 H_{K,in}$ is determined to be within 6 – 39 mT, which justifies the condition necessary for Eq. (1), *i.e.*, $H_{K,in} < H_{RKKY}$, and is consistent with the previous reports on the s-MTJs possessing strain-induced and/or magnetocrystalline anisotropy[18].

Now, we investigate the dependence of external-field robustness on junction size. Figure 3(b) shows the ratio of the relaxation times between the AP and P states, $\tau_{AP}/\tau_P$, as a function of $\mu_0 H_x$. In s-MTJs with uniaxial anisotropy, this ratio follows an exponential function, expressed as $\exp(A\mu_0 H_x)$, where $A = 2M_S V_m/(\mu_0 k_B T)$ represents the susceptibility to the applied field[17] as shown in Fig. 3(b). $M_S$ and $V_m$ are the spontaneous magnetization and the volume of magnetic layers, respectively. To compare with another susceptibility measure, we examine the slope of the sigmoidal response in the time-averaged behavior (inset in Fig. 3(b)), which is given by $s_H = M_S V_m / 2\mu_0 k_B T$[32], corresponding to $s_H = A/4$. Since the sigmoidal response is normalized to a range between 0 and +1, the inverse of the slope, $1/|s_H|$, gives a rough indication of the range of $H_x$ within which stable operation of the s-MTJ is expected.

Figure 3(c) shows $|s_H|$ as a function of $D_{ele}$. As $D_{ele}$ decreases, $|s_H|$ decreases, indicating that smaller s-MTJs exhibit greater robustness against external magnetic fields, again favorable for applications in terms of scalability. For $D_{ele}$ < 80 nm, $|s_H|$ is well below 0.1 mT$^{-1}$, meaning that stable operation is expected for magnetic field up to ~ 10 mT, which is larger than the typical value in practical environments[35]. In an ideal condition where the magnetization of the SAF-free layer is fully compensated, the relaxation time ratio $\tau_{AP}/\tau_P$ remains unity across the external field sweep (*i.e.*, $s_H$ = 0), whereas for SAF-free layer with an uncompensated magnetic moment, $\tau_{AP}/\tau_P$ deviates from unity due to finite Zeeman energy[17]. In SAF-free layer with uniaxial magnetic anisotropy, $s_H$ measured along the magnetic easy axis is given by a static-process calculation[32] as:

$$s_H = \frac{A}{4} \equiv \frac{1}{\mu_0} \frac{\partial(\tau_{AP}/\tau_P)}{\partial H_x}\bigg|_{\tau_{AP}/\tau_P=1} = \frac{\pi D^2 (m_2 - m_1)}{8\mu_0 k_B T}. \quad (2)$$

Assuming that $|s_H|$ is maximized when the easy axis is aligned along the +*x* direction and Zeeman energy is at its highest, fitting Eq. (2) to the experimental result yields the uncompensated magnetic moment $m_2-m_1$ in our s-MTJs to be 5% of total moment $m_2+m_1$.

We then turn to the dependence of stochastic behavior in SAF-free layer s-MTJs on the applied current *I* and voltage *V*. The time-averaged resistance ⟨*R*⟩ as a function of *I* is measured by the electrical circuit illustrated in Fig. 4(a). Figure 4(b) presents a representative ⟨*R*⟩–*I* curve under $\mu_0 H_x$ of ±100 mT and 0 T. When $\mu_0 H_x$ = +100 mT (−100 mT), the magnetization direction of the free layer aligns in the P (AP) state relative to the reference layer, while at 0 T, the averaged magnetization direction varies with the applied current. In our s-MTJs, the P state is energetically favored at zero field and zero current due to the stray field from the reference layer, whereas stochastic behavior is observed at 0 T by applying the positive bias current. To standardize the resistance, ⟨*R*⟩ at 0 T is normalized by ⟨*R*⟩ at the P (+100 mT) and AP (−100 mT) states, defining the normalized resistance as ⟨*r*⟩ ≡ (⟨*R*⟩ − ⟨*R*⟩$_P$)/(⟨*R*⟩$_{AP}$ − ⟨*R*⟩$_P$), as shown in Fig. 4(c). The sigmoidal function provides a good fit to $\langle r \rangle$, as depicted in the figure. We define parameter $s_I$ as $s_I \equiv \partial \langle r \rangle / \partial I |_{\langle r \rangle = 0.5}$, which quantifies the sensitivity of $\langle r \rangle$ to the bias current. Figure 4(d) summarizes $s_I$ measured at $\mu_0 H_x = 0$ T as a function of $D_{ele}$. In a previous study on s-MTJs having a single free layer with a perpendicular easy axis[32], $s_I$ gradually increased from 0.1 $(\mu A)^{-1}$ to 0.4 $(\mu A)^{-1}$ as $D_{ele}$ decreased from 60 nm to 25 nm. On the contrary, in Fig. 4(d), $s_I$ does not show any meaningful trend with respect to $D_{ele}$. It is also important to note that the magnitude of $s_I$ for in-plane s-MTJ is at least one order of magnitude smaller than that for perpendicular s-MTJs. This difference can be attributed to the difference in the spin-transfer torque (STT) efficiency between the in-plane and perpendicular MTJs[36], considering the fact that $s_I$ is analytically given by $\Delta/I_{C0}$[32], which is often used as a figure of merit of the STT switching efficiency relative to the energy barrier.

Figure 4(e) represents $s_V$ as a function of $D_{ele}$. We note that, for the stable operation of a probabilistic bit (p-bit) circuit[37], smaller $s_V$, *i.e.*, insensitive to bias voltage, is preferable[38]. $s_V$ appears to decrease with $D_{ele}$, following the relation $s_V \sim s_I/\langle R \rangle \propto D_{ele}^2$. To quantitatively assess the strength of this correlation, Pearson correlation coefficient $r$ and corresponding $p$-value are calculated under the assumption that $s_V \propto D_{ele}^2$. The results, $r = 0.77$ and $p$-value $= 0.0085$, indicate a statistically significant correlation between $s_V$ and $D_{ele}^2$. This is once again favorable for applications in terms of the scalability for the p-bit circuit[37]. At $D_{ele} \sim$ 40 nm, $s_V$ takes a value of approximately 0.5 V$^{-1}$. Previous studies on s-MTJs[32] with a perpendicular easy axis reported $s_I \approx 0.3$ $(\mu A)^{-1}$ at $D_{ele} = 40$ nm for devices with an $RA$ of 6.5 $\Omega \cdot \mu m^2$, corresponding to $s_V = 58$ V$^{-1}$. This discrepancy primarily arises from the difference in $s_I$ between in-plane and perpendicular s-MTJs and, subsequently, their respective $RA$ values. We also note that the bias voltage insensitivity can be enhanced by increasing $RA$ of s-MTJ as well; however, this potentially causes an increase in RC delay, requiring careful optimization.

The reliable and stable operation of s-MTJ-based p-bit[28,37] with respect to the voltage noise is evaluated by $s_V V_{DD,noise}/2$, where $V_{DD,noise}$ represents the power supply noise. Semiconductor industry standards and technical guidelines generally recommend maintaining $V_{DD}$ noise within a few percent of the supply voltage, corresponding to a range from a few millivolts to a few tens of millivolts, depending on the application and the sensitivity of the devices. For instance, JEDEC standards for DDR3 memory specify that reference voltage deviations within ±1% of $V_{DD}$ (~±15 mV)[39], while Intel FPGA design guidelines restrict core voltage variations to within ±30–50 mV[40]. Assuming $V_{DD,noise}$ = 50 mV, the obtained $s_V$ of 0.5 V$^{-1}$ for the current SAF free layer s-MTJ leads to the P/AP probability variation of 1.3%, which is reasonably small for most applications. This is contrasting with the case with perpendicular s-MTJs with $s_V$ = 58 V$^{-1}$, which corresponds to the variation of 140%.

In summary, we have systematically investigated the stochastic properties of synthetic antiferromagnetic free-layer stochastic magnetic tunnel junctions (SAF s-MTJs) with varying junction sizes under external perturbations such as magnetic fields and bias voltages. Our experimental results, supported by an analytical model, have revealed several insights/benefits: (1) smaller devices exhibit shorter relaxation times due to the reduced energy barriers; (2) robustness against external magnetic fields improves significantly with reduced junction size; (3) insensitivity to bias voltage improves in smaller junctions; (4) sensitivity to bias voltage is at least one order of magnitude smaller than that with the previously studied s-MTJ with perpendicular easy axis[32], primarily attributed to difference in spin-transfer torque efficiency. These findings provide comprehensive guidelines for designing highly robust and efficient s-MTJ devices with a SAF-free layer for spintronics-based probabilistic computing applications.

**Acknowledgments**

The authors thank F. Shibata, R. Nomura, K. Kino, I. Morita, R. Ono, and M. Musya for their technical support. This work was partly supported by JST-CREST (Grant No. JPMJCR19K3), JST-PRESTO (Grant No. JPMJPR21B2), JST-ASPIRE (Grant No. JPMJAP2322), MEXT Initiative to Establish Next-generation Novel Integrated Circuits Centers (X-NICS) (Grant No. JPJ011438), JSPS Kakenhi (Grant No. 24H02235, No. 24H00039 and 24K22949), and RIEC Cooperative Research Projects.

# References

1) B. Sutton, K.Y. Camsari, B. Behin-Aein, and S. Datta, Sci. Rep. **7**, 44370 (2017).

2) K.Y. Camsari, R. Faria, B.M. Sutton, and S. Datta, Phys. Rev. X **7**, 031014 (2017).

3) Y. Liu, Q. Hu, Q. Wu, X. Liu, Y. Zhao, D. Zhang, Z. Han, J. Cheng, Q. Ding, Y. Han, B. Peng, H. Jiang, X. Xue, H. Lv, and J. Yang, Micromachines **13**, 924 (2022).

4) K.S. Woo, J. Kim, J. Han, W. Kim, Y.H. Jang, and C.S. Hwang, Nat. Commun. **13**, 5762 (2022).

5) T.J. Park, K. Selcuk, H.-T. Zhang, S. Manna, R. Batra, Q. Wang, H. Yu, N.A. Aadit, S.K.R.S. Sankaranarayanan, H. Zhou, K.Y. Camsari, and S. Ramanathan, Nano Lett. **22**, 8654 (2022).

6) G.M. Gutiérrez-Finol, S. Giménez-Santamarina, Z. Hu, L.E. Rosaleny, S. Cardona-Serra, and A. Gaita-Ariño, Npj Comput. Mater. **9**, 1 (2023).

7) W. Whitehead, Z. Nelson, K.Y. Camsari, and L. Theogarajan, Nat. Electron. **6**, 1009 (2023).

8) S. Luo, Y. He, B. Cai, X. Gong, and G. Liang, IEEE Electron Device Lett. **44**, 1356 (2023).

9) J. Wu, H. Sun, and G. Zhou, Small **20**, 2403755 (2024).

10) S. Chowdhury, A. Grimaldi, N.A. Aadit, S. Niazi, M. Mohseni, S. Kanai, H. Ohno, S. Fukami, L. Theogarajan, G. Finocchio, S. Datta, and K.Y. Camsari, IEEE J. Explor. Solid-State Comput. Devices Circuits **9**, 1 (2023).

11) D. Vodenicarevic, N. Locatelli, A. Mizrahi, J.S. Friedman, A.F. Vincent, M. Romera, A. Fukushima, K. Yakushiji, H. Kubota, S. Yuasa, S. Tiwari, J. Grollier, and D. Querlioz, Phys. Rev. Appl. **8**, 054045 (2017).

12) B. Parks, M. Bapna, J. Igbokwe, H. Almasi, W. Wang, and S.A. Majetich, AIP Adv. **8**, 055903 (2017).

13) S. Kanai, K. Hayakawa, H. Ohno, and S. Fukami, Phys. Rev. B **103**, 094423 (2021).

14) K. Hayakawa, S. Kanai, T. Funatsu, J. Igarashi, B. Jinnai, W.A. Borders, H. Ohno, and S. Fukami, Phys. Rev. Lett. **126**, 117202 (2021).

15) K.Y. Camsari, M.M. Torunbalci, W.A. Borders, H. Ohno, and S. Fukami, Phys. Rev. Appl. **15**, 044049 (2021).

16) C. Safranski, J. Kaiser, P. Trouilloud, P. Hashemi, G. Hu, and J.Z. Sun, Nano Lett. **21**, 2040 (2021).

17) K. Kobayashi, K. Hayakawa, J. Igarashi, W.A. Borders, S. Kanai, H. Ohno, and S. Fukami, Phys. Rev. Appl. **18**, 054085 (2022).

18) J.Z. Sun, C. Safranski, P. Trouilloud, C. D'Emic, P. Hashemi, and G. Hu, Phys. Rev. B **107**, 184433 (2023).

19) J.Z. Sun, C. Safranski, P. Trouilloud, C. D'Emic, P. Hashemi, and G. Hu, Phys. Rev. B **108**, 064418 (2023).

20) K. Selcuk, S. Kanai, R. Ota, H. Ohno, S. Fukami, and K.Y. Camsari, Phys. Rev. Appl. **21**, 054002 (2024).

21) R. Ota, K. Kobayashi, K. Hayakawa, S. Kanai, K.Y. Çamsarı, H. Ohno, and S. Fukami, Appl. Phys. Lett. **125**, 022406 (2024).

22) A. Mizrahi, T. Hirtzlin, A. Fukushima, H. Kubota, S. Yuasa, J. Grollier, and D. Querlioz, Nat. Commun. **9**, 1533 (2018).

23) W.A. Borders, A.Z. Pervaiz, S. Fukami, K.Y. Camsari, H. Ohno, and S. Datta, Nature **573**, 390 (2019).

24) Y. Shao, S.L. Sinaga, I.O. Sunmola, A.S. Borland, M.J. Carey, J.A. Katine, V. Lopez-Dominguez, and P.K. Amiri, IEEE Magn. Lett. **12**, 9395199 (2021).

25) J. Kaiser, W.A. Borders, K.Y. Camsari, S. Fukami, H. Ohno, and S. Datta, Phys. Rev. Appl. **17**, 014016 (2022).

26) A. Grimaldi, K. Selcuk, N.A. Aadit, K. Kobayashi, Q. Cao, S. Chowdhury, G. Finocchio, S. Kanai, H. Ohno, S. Fukami, and K.Y. Camsari, Int. Electron Devices Meet. 2022, 10.1109/ IEDM45625.2022.10019530.

27) N.S. Singh, S. Niazi, S. Chowdhury, K. Selcuk, H. Kaneko, K. Kobayashi, S. Kanai, H. Ohno, S. Fukami, and K.Y. Camsari, Int. Electron Devices Meet. 2023, 10.1109/ IEDM45741.2023.10413686.

28) N.S. Singh, K. Kobayashi, Q. Cao, K. Selcuk, T. Hu, S. Niazi, N.A. Aadit, S. Kanai, H. Ohno, S. Fukami, and K.Y. Camsari, Nat. Commun. **15**, 2685 (2024).

29) J. Si, S. Yang, Y. Cen, J. Chen, Y. Huang, Z. Yao, D.-J. Kim, K. Cai, J. Yoo, X. Fong, and H. Yang, Nat. Commun. **15**, 3457 (2024).

30) N.S. Singh, C. Delacour, S. Niazi, K. Selcuk, D. Golenchenko, H. Kaneko, S. Kanai, H. Ohno, S. Fukami, and K.Y. Camsari, Int. Electron Devices Meet. 2024, 10.1109/ IEDM50854.2024.10873478.

31) D. Chen, A.D. Kent, D. Sels, and F. Morone, Phys. Rev. Res. **7**, 013129 (2025).

32) K. Kobayashi, W.A. Borders, S. Kanai, K. Hayakawa, H. Ohno, and S. Fukami, Appl. Phys. Lett. **119**, 132406 (2021).

33) D. C. Worledge, Appl. Phys. Lett. **84**, 2847 (2004).

34) D. C. Worledge, Appl. Phys. Lett. **84**, 4559 (2004).

35) B. Dieny, S. Aggarwal, V.B. Naik, S. Couet, T. Coughlin, S. Fukami, K. Garello, J. Guedj, J.A.C. Incorvia, L. Lebrun, K.-J. Lee, D. Leonelli, Y. Noh, S. Salimy, S. Soss, L. Thomas, W. Wang, and D.C. Worledge, IEEE Electron Devices Mag. **2**, 52 (2024).

36) S. Mangin, D. Ravelosona, J.A. Katine, M.J. Carey, B.D. Terris, and E.E. Fullerton, Nat. Mater. **5**, 210 (2006).

37) K.Y. Camsari, S. Salahuddin, and S. Datta, IEEE Electron Device Lett. **38**, 1767 (2017).

38) O. Hassan, S. Datta, and K.Y. Camsari, Phys. Rev. Appl. **15**, 064046 (2021).

39) JEDEC Solid State Technology Association, "DDR3 SDRAM Standard JESD79-3F," Arlington, VA, USA (2012). Available: https://www.jedec.org/standards-documents

40) Intel Corporation, "Cyclone 10 LP Device Pin Connection Guidelines," Document AN-800, Intel FPGA, San Jose, CA, USA (2017). Available: https://www.intel.com/content/www/us/en/docs/programmable/683458/current/cyclone-10-lp-device-pin-connection.html

**Figures**

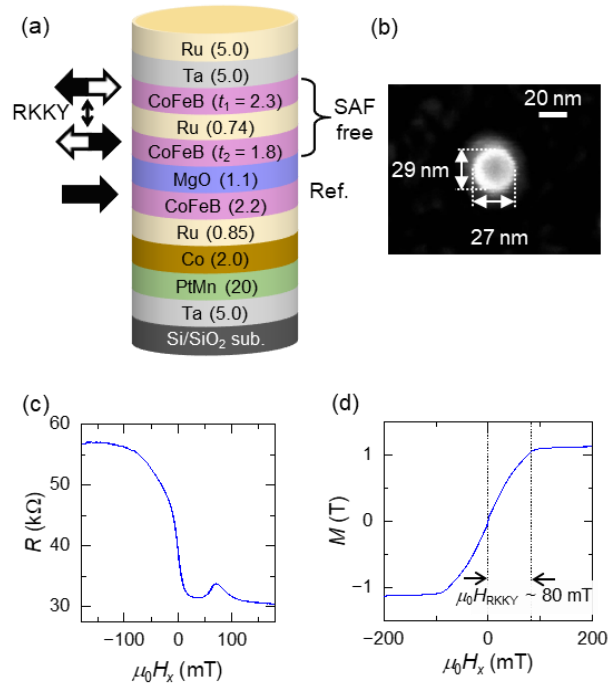

**Fig. 1.** (a) Stack structure of stochastic magnetic tunnel junction (s-MTJ) with a synthetic antiferromagnetic (SAF) free layer. (b) Scanning electron microscope image of s-MTJ. (c) Resistance $R$ of a s-MTJ with a SAF free layer as a function of $\mu_0 H_x$, where $\mu_0$ is the permeability of the vacuum. (d) Magnetization $M$ as a function of in-plane magnetic field $\mu_0 H_x$ for a blanket film with identical free layer structure as (a).

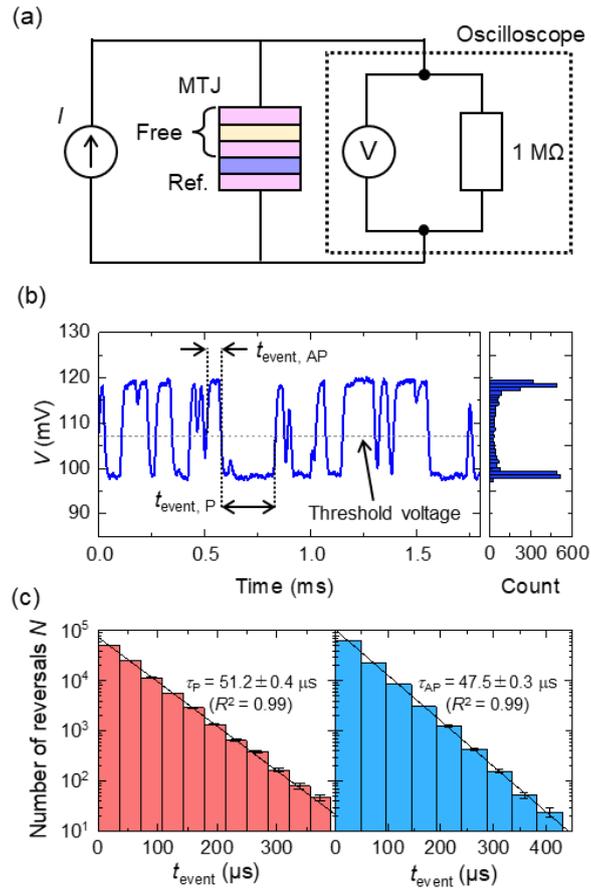

**Fig. 2.** (a) Electrical circuit for the random telegraph noise (RTN) measurements. (b) RTN signal of s-MTJ under an applied current of 2.4 µA, with definitions of event times $t_{event}$ along with a histogram showing the distribution of voltage levels $V$. (d) Histogram of number, $N$, of magnetization reversals as a function of $t_{event}$.

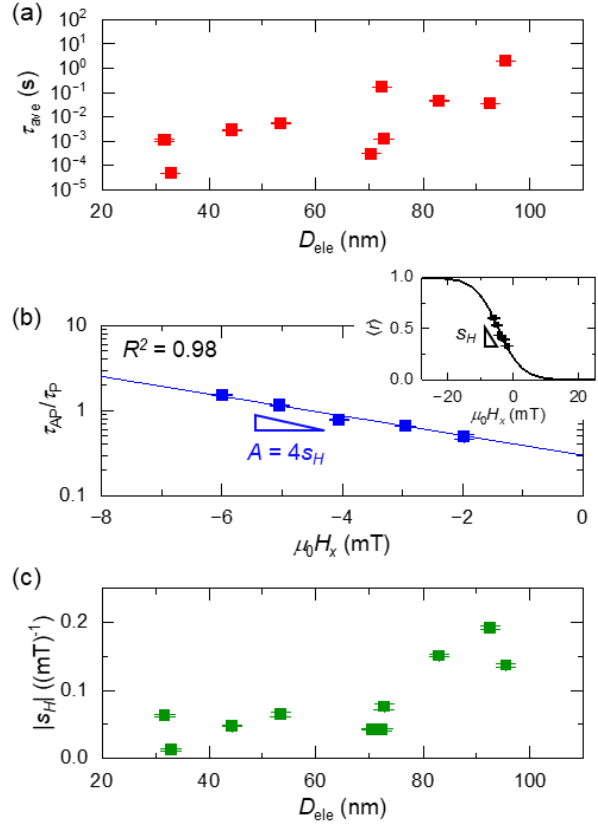

**Fig. 3.** (a) Relaxation time as a function of electrically determined diameter $D_{ele}$ of the s-MTJs averaged for P and AP states. (b) $\tau_{AP}/\tau_P$ as a function of external in-plane magnetic field $\mu_0 H_x$ and definition of slope $A$. Inset: Normalized time-averaged resistance $\langle r \rangle$ as a function of $\mu_0 H_x$ and definition of slope $s_H$. (c) $|s_H|$ as a function of $D_{ele}$.

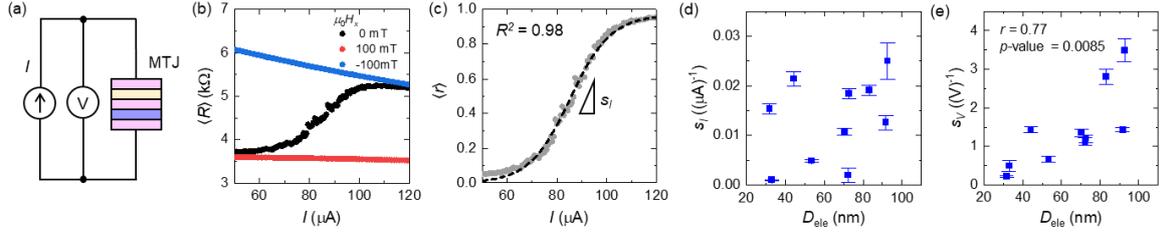

**Fig. 4.** (a) Electrical circuit for ⟨$R$⟩-$I$ curve measurements. (b) An example of ⟨$R$⟩-$I$ curves under various magnetic fields. The red (blue) plots with the applied field of 100 mT (-100 mT) represent the ⟨$R$⟩ of P (AP) state. The integration time for each point is set so that more than 100 magnetization reversals are contained. (c) Normalized ⟨$R$⟩ as a function of applied current and definition of $s_I$. Normalized ⟨$R$⟩ = 1 (0) corresponds to the AP (P) state of the MTJ. Black dashed line indicates the fitting result with a sigmoidal function. (d) $s_I$ as a function of $D_{ele}$. (e) $s_V$ as a function of $D_{ele}$.